\documentclass[review]{elsarticle}

\usepackage{lineno,hyperref}

\usepackage[margin=1in]{geometry}

\pdfoutput=1

\usepackage{graphicx}
\usepackage{bm}
\usepackage{dcolumn}
\usepackage{setspace}
\usepackage{amsmath}
\usepackage{fancyhdr}

\usepackage{caption}
\usepackage{subcaption}

\usepackage{color,soul}

\renewcommand\hl[1]{#1}

\modulolinenumbers[5]

\begin{document}

\begin{frontmatter}

\title{Role of twin boundary position on the yield strength of \hl{Cu} nanopillars}

%% Group authors per affiliation: %\author{Elsevier\fnref{myfootnote}} %\address{Radarweg 29, Amsterdam} %\fntext[myfootnote]{Since 1880.}

\author[MDTD]{G. Sainath\corref{mycorrespondingauthor}}
\cortext[mycorrespondingauthor]{Corresponding author}
\ead{sg@igcar.gov.in}

\author[MDTD,HBNI,Technion]{P. Rohith}

\author[MDTD,HBNI]{A. Nagesha}

\address[MDTD]{Materials Development and Technology Division, Metallurgy and Materials Group, Indira Gandhi Centre for Atomic Research, Kalpakkam, Tamilnadu-603102, India}

\address[HBNI]{Homi Bhabha National Institute, Indira Gandhi Centre for Atomic Research, Kalpakkam, Tamilnadu-603102, India}

\address[Technion]{Present address, Faculty of Mechanical Engineering, Technion Israel Institute of Technology, Haifa - 32000, Haifa, Israel}

\begin{abstract}

It is well known that the twin boundary (TB) spacing plays an important role in controlling the strength of twinned 
metallic nanopillars. One of the reasons attributed to this strengthening behaviour is the force exerted by the TBs 
on dislocations. Since the TBs exert repulsive force on dislocations and the plasticity in nanopillars is surface 
controlled, it is interesting to know whether the TB position from the nanowire surface has any effect on the 
strength of twinned nanopillars. Using atomistic simulations, here we show that the TB position significantly 
influences the strength of twinned nanopillars. Atomistic simulations have been performed on nanopillar containing 
one and two TBs and their position is varied within nanopillar from the center to the surface. The results indicate 
that in nanopillar containing a single TB, the strength is maximum when the TB is located at the center of the 
nanopillar and it decreases as the TB is shifted towards the nanopillar surface. On the other hand in nanopillar 
containing two TBs, the maximum strength is observed when the twin boundaries are placed at distance of one fourth 
\hl{or one fifth} the pillar size from the surfaces and it decreases when TBs are moved on either side. The present 
study demonstrates that the mechanical properties of the twinned nanopillars can be controlled by carefully tailoring 
the position of the TBs within the nanopillars.
\end{abstract}


\end{frontmatter}

%\linenumbers

\section{Introduction}
In recent years, the mechanical properties and associated plastic deformation mechanisms in nanowires/ nanopillars 
have received extensive research interest by means of in-situ experiments and atomistic simulations. Deformation 
behavior in these nano systems beyond their elastic limit is found to be quite different from their bulk 
counterparts \cite{Review1-Cai,Review2-EML}. In nanopillars/nanowires, the surface mediated plasticity has been 
identified as a dominant deformation mechanism \cite{Review2-EML,PRL-Samanta,Volkert-APL}, while in bulk materials 
dislocation multiplication, pile-up, cross-slip and other forest mechanisms are known to play the dominant role. 
The yielding in FCC nanopillars occurs through the nucleation of Shockley partials or extended dislocations from 
the surface and it can be influenced by a variety of factors like nanowire size, shape, orientation, temperature 
and strain rate \cite{Review1-Cai,PRL-Samanta,Cao2008-Shape,Xie2015-Srate,Rohith-CCM,Sainath-PhilMag17}. 
As a result, the yield strength of the nanopillars is also sensitive to all these factors.

In addition to the above mentioned factors, the presence of planar defects like twin boundaries (TBs) also influences 
the yield strength and deformation behavior of nanopillars. The introduction of TBs has become a well-known means of 
strengthening method in nanopillars and nanocrystalline materials. As a result, many experimental and atomistic 
simulation studies have been performed on twinned nanopillars containing high density of TBs 
\cite{Cao2007-TBs,Sansoz-NanoLett,Lu2009-Sci,Jang2012-Natnano,Rohith-PhilMag19}. These studies have revealed that the 
introduction of TBs increases the strength along with ductility of twinned nanopillars as compared to their twin free 
counterparts. The dual nature of TBs to acts as an obstacle as well as to aid the dislocation motion results in 
increasing the strength without the loss of ductility \cite{Cao2007-TBs}. It has also been found that the strengthening 
effect of TBs strongly depends on the TB spacing \cite{Cao2007-TBs,Sansoz-NanoLett,Yang2017-SciRep,Sainath-PhilMag16}. 
Generally, the yield strength of twinned nanopillars increases with decreasing twin boundary (TB) spacing 
\cite{Sansoz-NanoLett,Rohith-PhilMag19,Yang2017-SciRep} resembling the Hall-Petch type behaviour in polycrystalline 
materials. This strengthening effect in twinned nanopillars has been attributed  to the additional repulsive force 
exerted by TBs on dislocation nucleation and glide \cite{RF-PRB,RF-Sansoz,RF-Acta}.

In addition to TB spacing, in longitudinally twinned nanopillars, where TBs run parallel to pillar surfaces, it is 
interesting to investigate the role of TB position from the nanopillar surfaces. It becomes even more important since 
the plasticity in nanopillars is surface controlled \cite{Review2-EML,PRL-Samanta,Volkert-APL} and TBs exerts repulsive 
force on dislocation nucleation \cite{RF-PRB,RF-Sansoz,RF-Acta}. Due to limited studies on twinned nanopillars containing 
longitudinal TBs \cite{Rohith-PhilMag19,SainathPLA,PRL-TBP}, the role of TB position has not attracted the attention of 
researchers, especially on yield strength. A recent study by Cheng and co-workers \cite{PRL-TBP} has shown that when the 
TB is placed close to the nanopillar surface, the deformation dominated by anomalous de-twinning mechanism completely 
annihilate the TB and leads to the formation of a single crystal. This annihilation of longitudinal TB has not been 
observed when the TB is placed at the center of the pillar \cite{SainathPLA,Acta-Axial,Jeon-Scripta}. These results 
suggest that the TB position play an important role in longitudinally twinned nanopillars. In view of this, the present 
study is aimed at understanding the effect of TB position on the strength of Cu nanopillars using atomistic simulations. 
The $<$110$>$ and $<$112$>$ oriented Cu nanopillars containing one and two TBs running parallel to the axial direction 
were considered for this study. The position of the TB is varied from the surface to center of the nanopillar. 

\section{Simulation Methodology}

\begin{figure}
\centering
\includegraphics[width=7cm]{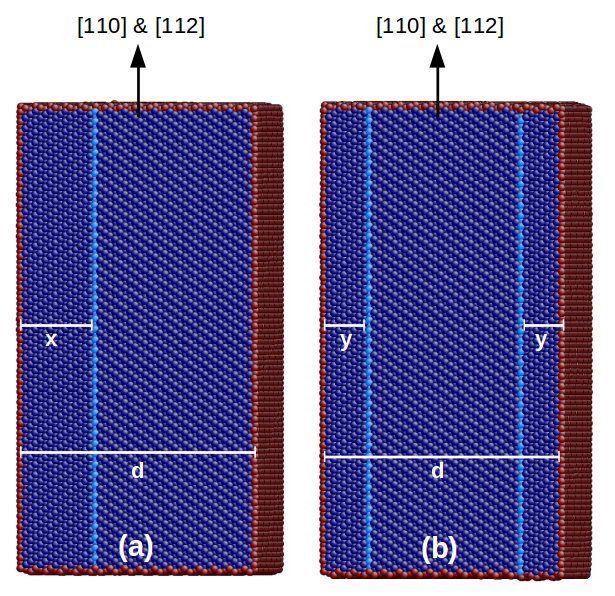}
\caption{The initial atomic configuration of twinned Cu nanopillar containing, (a) single twin boundary and (b) two twin 
boundaries, running parallel to the nanopillar axis ($<$110$>$ and $<$112$>$). In these nanopillars, the twin boundary 
position has been varied by varying x (in case of single twin boundary) and y (in case of two twin boundaries). The 
atoms are coloured according to their CNA parameter. The blue colour atoms indicate pure FCC Cu atoms, while cyan colour 
atoms represent the HCP or twin boundary atoms and the red colour atoms represent the surfaces.}
\label{Initial}
\end{figure}

Twinned Cu nanopillars with an axial orientation of $<$110$>$ and $<$112$>$ direction have been considered in this study. 
The $<$110$>$ nanopillar is enclosed by \{111\} and \{112\} type side surfaces, while the $<$112$>$ nanopillar has \{111\} 
and \{110\} side surfaces. In these two nanopillars, one and two TBs have been introduced parallel to the nanopillar axis 
and their positions or distance (x and y) is varied with respect to nanopillar surface as shown in Figure \ref{Initial}a-b. 
In order to examine the size effects, twinned nanopillars with two different cross-section widths (d) = 10 and 15 nm 
have been investigated in this study. The nanopillar length (l) was twice the cross-section width (d). On these twinned 
nanopillars, the tensile loading has been simulated using molecular dynamics (MD) simulations. All MD simulations were 
carried out in LAMMPS package \cite{Plimpton-1995} employing an an embedded atom method (EAM) potential for Cu given by 
Mishin and co-workers \cite{Mishin-2001}. This potential has been chosen for being able to reproduce stacking fault and 
twinning fault energies for Cu \cite{Liang-PRB}, which are key variables for predicting the dislocation nucleation and 
deformation mechanism related studies.

To mimic the geometry of nanopillars, no periodic boundary conditions were used in any direction during the simulations. 
Before applying tensile load, energy minimization was performed by conjugate gradient method to obtain a relaxed structure. 
To obtain the trajectory of atoms, velocity verlet algorithm was used to integrate the equations of motion with a time step 
of 2 fs. Following the energy minimization, the twinned nanopillars were equilibrated to required temperature of 10 K in a 
canonical ensemble (constant NVT) with a Nose-Hoover thermostat. Upon thermal equilibration, the deformation was carried out 
in a displacement-controlled manner at constant strain rate of $1 \times 10^8$ s$^{-1}$ by imposing displacements to atoms 
along the nanowire length that varied linearly from zero at the bottom fixed layer to a maximum value at the top fixed layer. 
The strain rate considered in this study is significantly higher than the experimental strain rates, nevertheless, it's in 
the conventional range of usual MD simulations. The strain ($\varepsilon$) was calculated as $(l-l_0)/l_0$, where $l$ 
is instantaneous length and $l_0$ is the initial length of the nanowire. The stress was obtained using the Virial 
expression \cite{Virial}, which is equivalent to a Cauchy's stress in an average sense. AtomEye package \cite{AtomEye}and 
OVITO \cite{Ovito} have been used for the visualization of atomic snapshots with common neighbour analysis (CNA).

\section{Results and Discussion}

%\subsection{Nanopillar with single twin boundary}

\begin{figure}[h]
\centering
\begin{subfigure}[b]{0.495\textwidth}
\includegraphics[width=\textwidth]{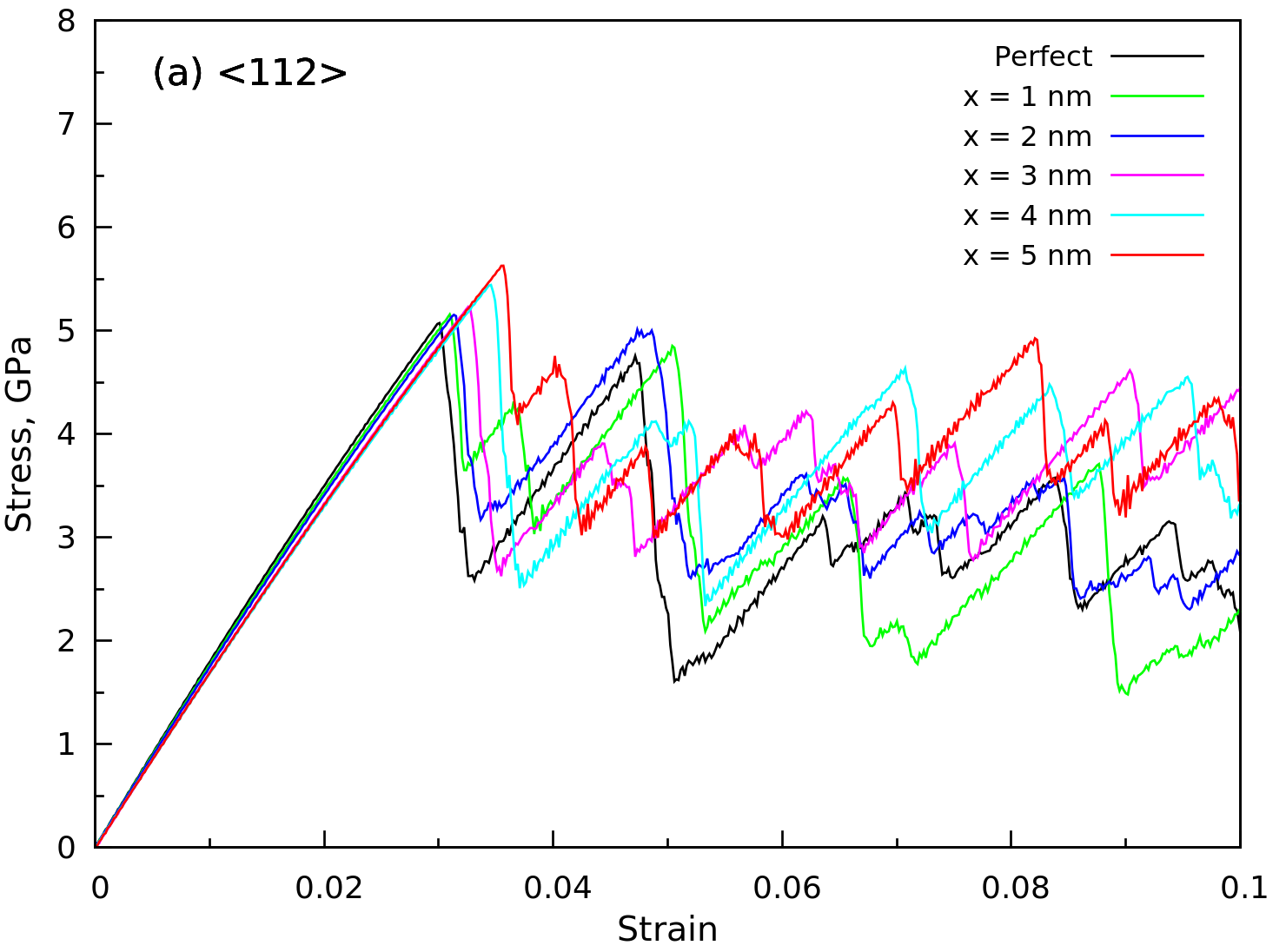}
%                \caption{}
\end{subfigure}
%\qquad
\begin{subfigure}[b]{0.475\textwidth}
\includegraphics[width=\textwidth]{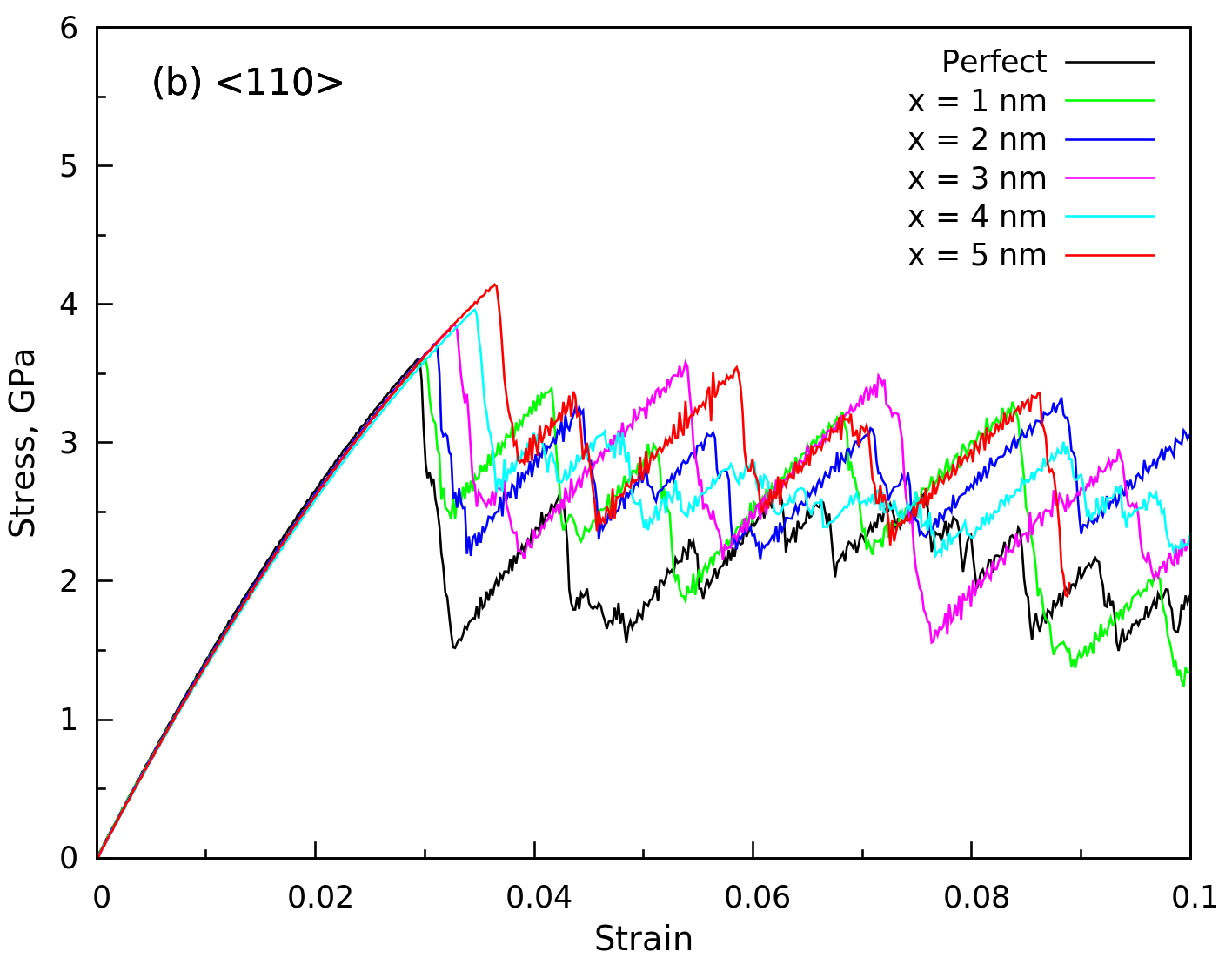}
%  \caption{}
\end{subfigure}
 \caption {The stress-strain behaviour of (a) $<$112$>$ and (b) $<$110$>$ oriented Cu nanopillars with d = 10 nm containing a 
 single twin boundary located at various distances (x) from the nearby by surface.} 
 \label{stress-strain}
 \end{figure}

\begin{figure}
\centering
\includegraphics[width=8cm]{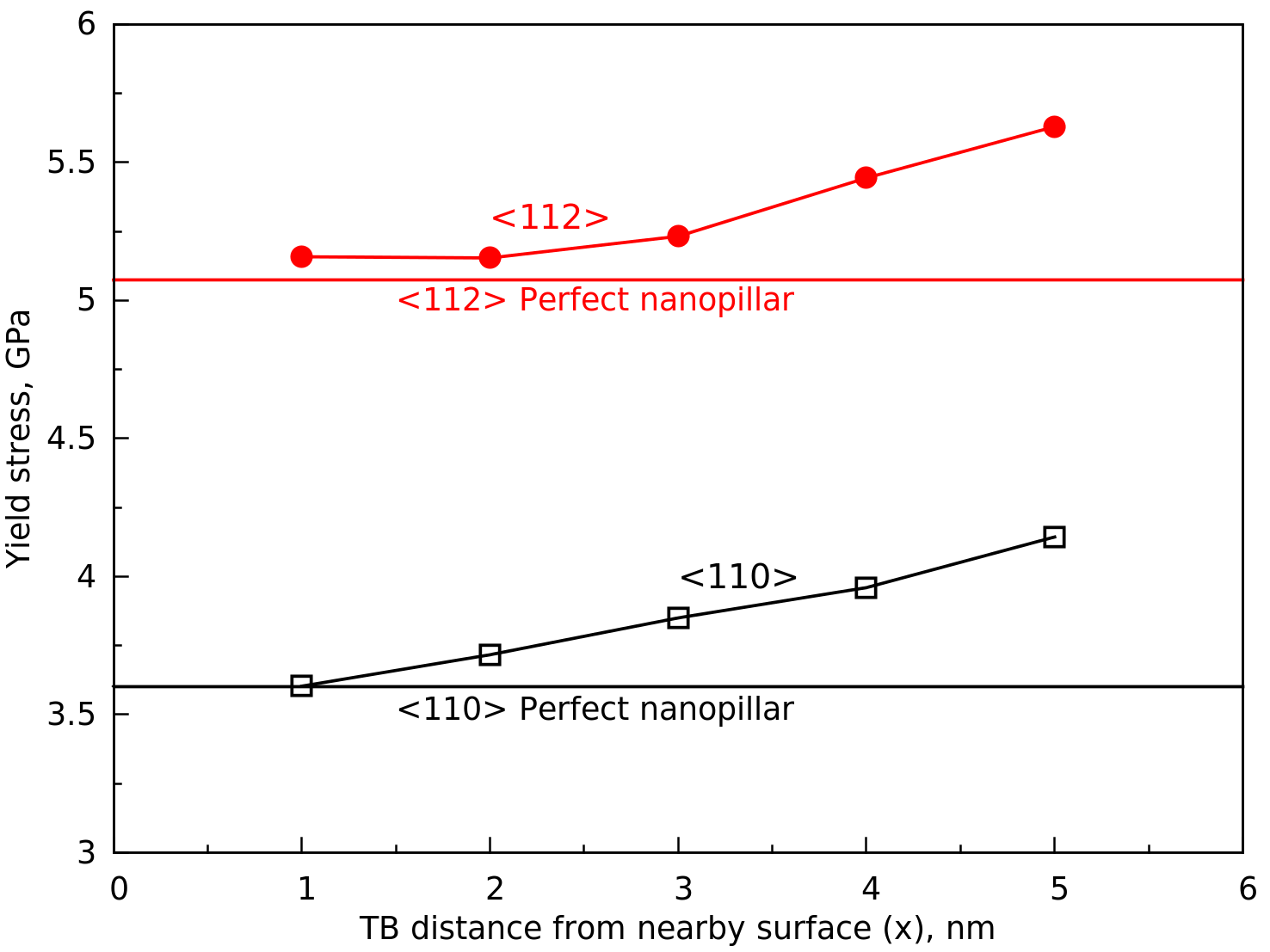}
\caption {The variation of yield strength in $<$112$>$ and $<$110$>$ twinned Cu nanopillars with d = 10 nm containing a single 
twin as a function of TB distance (x) from the nearby pillar surface. For comparison, the yield strength of perfect nanopillars 
in the respective orientations is also shown as horizontal line.}
\label{yieldstress}
\end{figure}

Figure \ref{stress-strain}(a)-(b) shows the tensile stress-strain behaviour of $<$112$>$ and $<$110$>$ oriented twinned Cu nanopillars 
containing a single axial TB located at various positions or distance (x) from the surface of the nanopillar with $d = 10$ nm.
It can be seen that both the nanopillars, irrespective of TB position, exhibit an initial linear elastic deformation till the 
peak value followed by a sudden drop in flow stress. This sudden drop indicates the commencement of plastic deformation in the 
nanopillars. With further increase in strain, the flow stress shows large fluctuations consisting of peaks and troughs, a typical 
signature of discrete plastic deformation in nanopillars\cite{descrete}. Further in both the nanopillars ($<$110$>$ and $<$112$>$), 
it can be seen that the TB position has no effect on the elastic deformation (Figure \ref{stress-strain}). However, the peak stress 
at the end of elastic 
deformation known as yield stress shows a significant and systematic variation with respect to TB position. Figure \ref{yieldstress} 
depicts the variation of yield strength in $<$112$>$ and $<$110$>$ nanopillars as a function of TB distance (x) from the nearby 
pillar surface. For comparison, the yield strength of perfect nanopillars is also shown as horizontal line. It shows that in both 
the orientations, the yield strength increases with increasing the TB distance from the nanopillar surface (Figure \ref{yieldstress}). 
In other words, the strength is lowest when the TB is placed close to the surface and reaches maximum value when it is located at 
the center of the pillar (i.e., x = 5 nm), which is equal distance from both the surfaces. Further when the TB is close to 
surface (i.e., x = 1 nm), the strengthening effect of TB is negligible as the strength is almost comparable to the perfect 
nanopillar (Figure \ref{yieldstress}). Like twinned nanopillar with d = 10 nm, the yield strength of a nanopillar with d = 15 nm 
has also shown similar behaviour, however, the peak strength is observed at x = 7.5 nm. Finally in all the cases the strength of 
twinned nanopillars is always higher than the perfect nanopillar, as observed in previous investigations \cite{SainathPLA,Deng2009}.

\begin{figure}
\centering
\includegraphics[width=12cm]{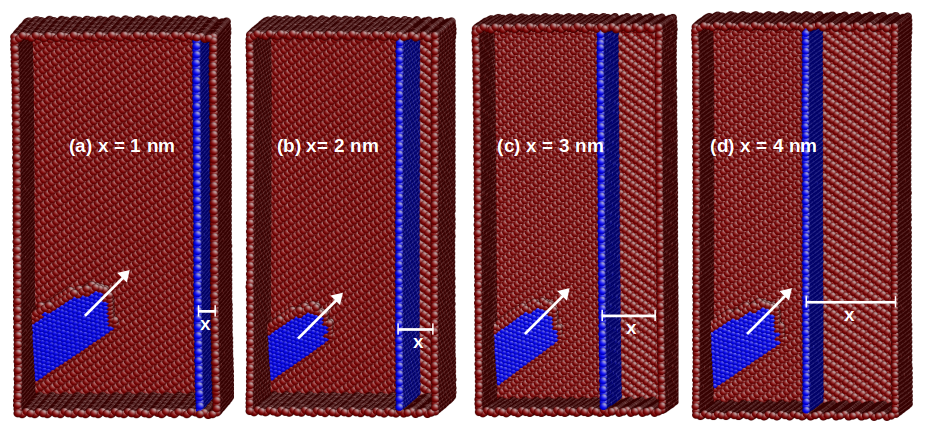}
\caption {The dislocation nucleation during the yielding of $<$112$>$ Cu nanopillar with d = 10 nm containing a single twin 
boundary located at a distance of (a) x = 1 nm, (b) x = 2 nm, (c) x = 3 nm, and (d) x = 4 nm from nanopillar surface. The atoms 
are coloured according to their CNA parameter. The front surfaces and the perfect atoms are removed for better visualization 
of dislocations in side the nanopillar. The blue colour atoms indicate the HCP or stacking fault atoms, while the red colour 
atoms represent the surfaces along with dislocation core atoms.}
\label{yielding}
\end{figure}

\begin{figure}[h]
\centering
\includegraphics[width=9cm]{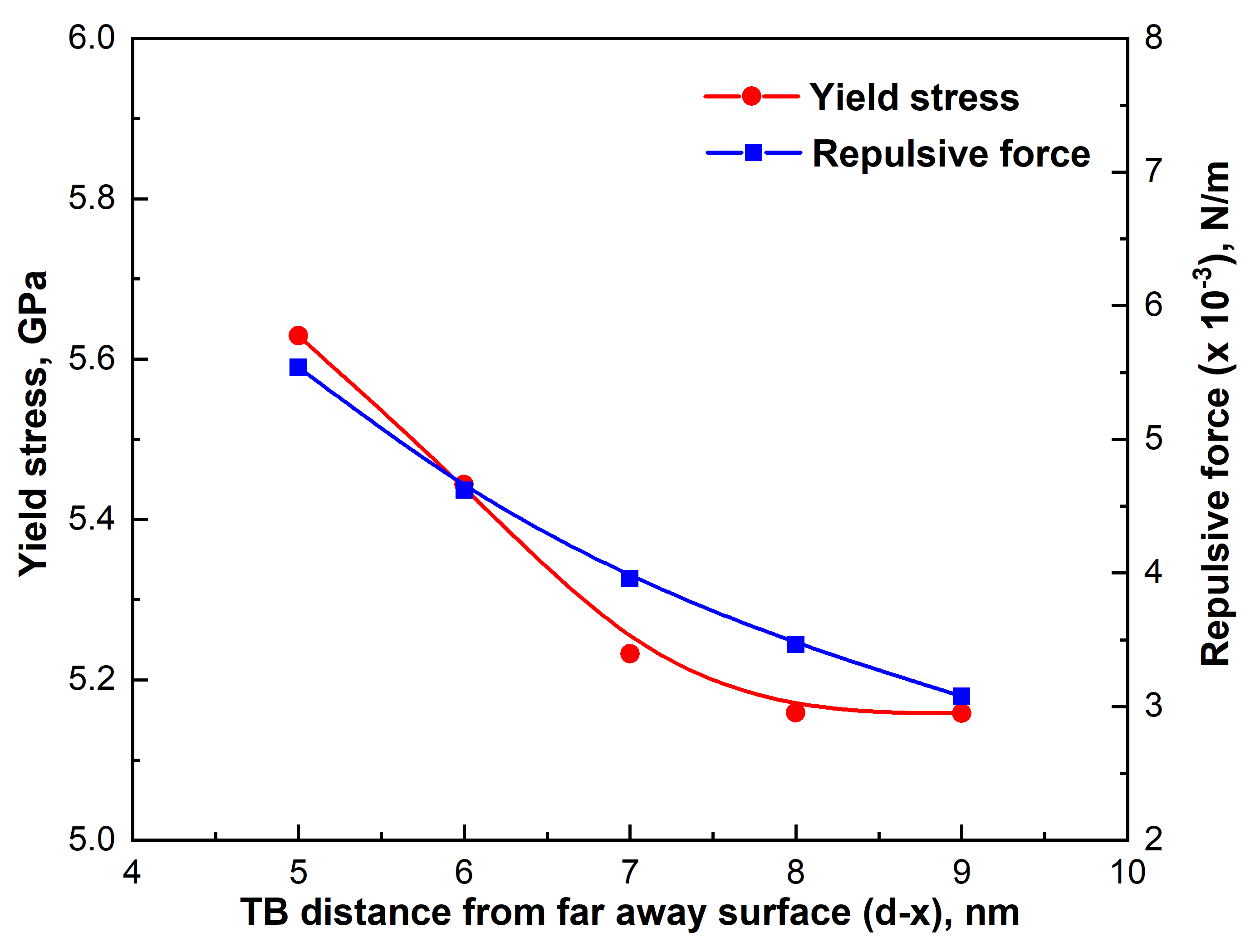}
\caption {The variation of repulsive force and the yield strength of $<$112$>$ twinned Cu nanopillar containing a single twin 
as a function of TB distance from the far away surface (d-x).}
\label{RF-YS}
\end{figure}

In twinned nanopillars the increase in strength has been attributed to the repulsive force offered by the TBs on dislocation 
nucleation \cite{RF-PRB,RF-Sansoz,RF-Acta}. The repulsive force ($f_{tb}$) exerted by the twin boundary on dislocations nucleating 
from a surface at a distance of $x$ is given by \cite{RF-PRB,RF-Sansoz,RF-Acta}
\begin{equation}
\centering
f_{tb} = \frac{\lambda\mu b^2}{4\pi x}
\label{equation1}
\end{equation} Here, $\lambda$ is dimensionless measure of the interaction strength due to the elasticity mismatch between 
the twin and perfect crystal, $\mu$ is the shear modulus in the slip direction and b is the Burgers vector of partial 
dislocations. According to this equation, the strength should be maximum when the TB is close to nanopillar surface. However, 
the reverse behaviour is observed in Figure \ref{yieldstress}. In order to understand this peculiar behaviour, the atomic 
configurations of nanopillars during the yielding have been captured and analysed. Figure \ref{yielding} shows the yielding 
in Cu nanopillar having a single TB at various distance from pillar surface. It can be seen that at all TB positions, the 
yielding occurs through the nucleation of 1/6$<$112$>$ Shockley partial dislocations from the nanopillar corners. Further, 
the nucleation occurs only in the larger grain, i.e, from corners that are far away from the TB (Figure \ref{yielding}). No 
nucleation of dislocations is observed in the smaller grain. As the Shockley partials nucleate from a surface which is far 
away from the TB, the distance between TB position and far away surface has to be considered in repulsive force calculations. 
In a nanopillar with single TB located at a distance of x nm from the nearby surface, the same boundary is located at a 
distance of ``d-x'' nm away from the far away surface, where d is the cross-section width of the nanopillar. Accordingly, 
the repulsive force equation has to be modified as follows; 
\begin{equation}
\centering
f_{tb} = \frac{\lambda\mu b^2}{4\pi(d-x)}
\label{equation}
\end{equation}Using this equation, the values of repulsive force have been calculated by substituting the appropriate values 
of $\lambda$, $\mu$ and $b$. For Cu, $\lambda$ = 0.3 \cite{RF-Sansoz} and $\mu$ = 53.31 GPa. The variations in repulsive 
force as a function of distance ``d-x'' is shown in Figure \ref{RF-YS}. For comparison, the yield stress values of $<$112$>$ 
twinned nanopillar were also plotted with respect to ``d-x''. It can be seen that, both yield stress and the repulsive force 
exhibit a similar decreasing behaviour with increasing distance ``d-x'' from the far away surface. Thus, there is clear 
correlation between the yield stress and the repulsive force on dislocations nucleating from the far away surface. This 
indicates that the distance between the TB and nanopillar surface dictate the strength of a nanopillar. 

%\subsection{Nanopillar with two twin boundaries}

\begin{figure}
\centering
\begin{subfigure}[b]{0.49\textwidth}
\includegraphics[width=\textwidth]{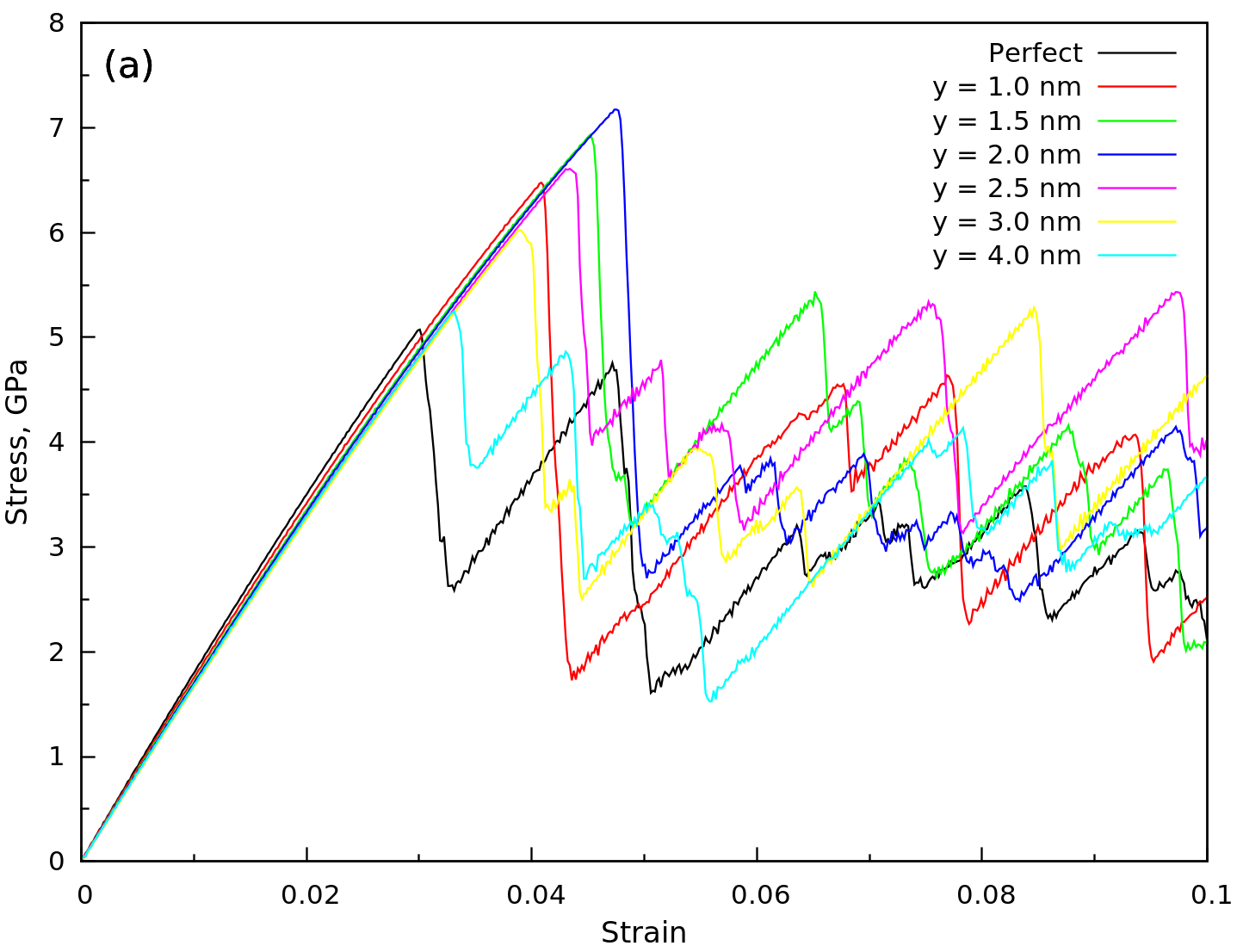}
%                \caption{}
\end{subfigure}
%\qquad
\begin{subfigure}[b]{0.48\textwidth}
\includegraphics[width=\textwidth]{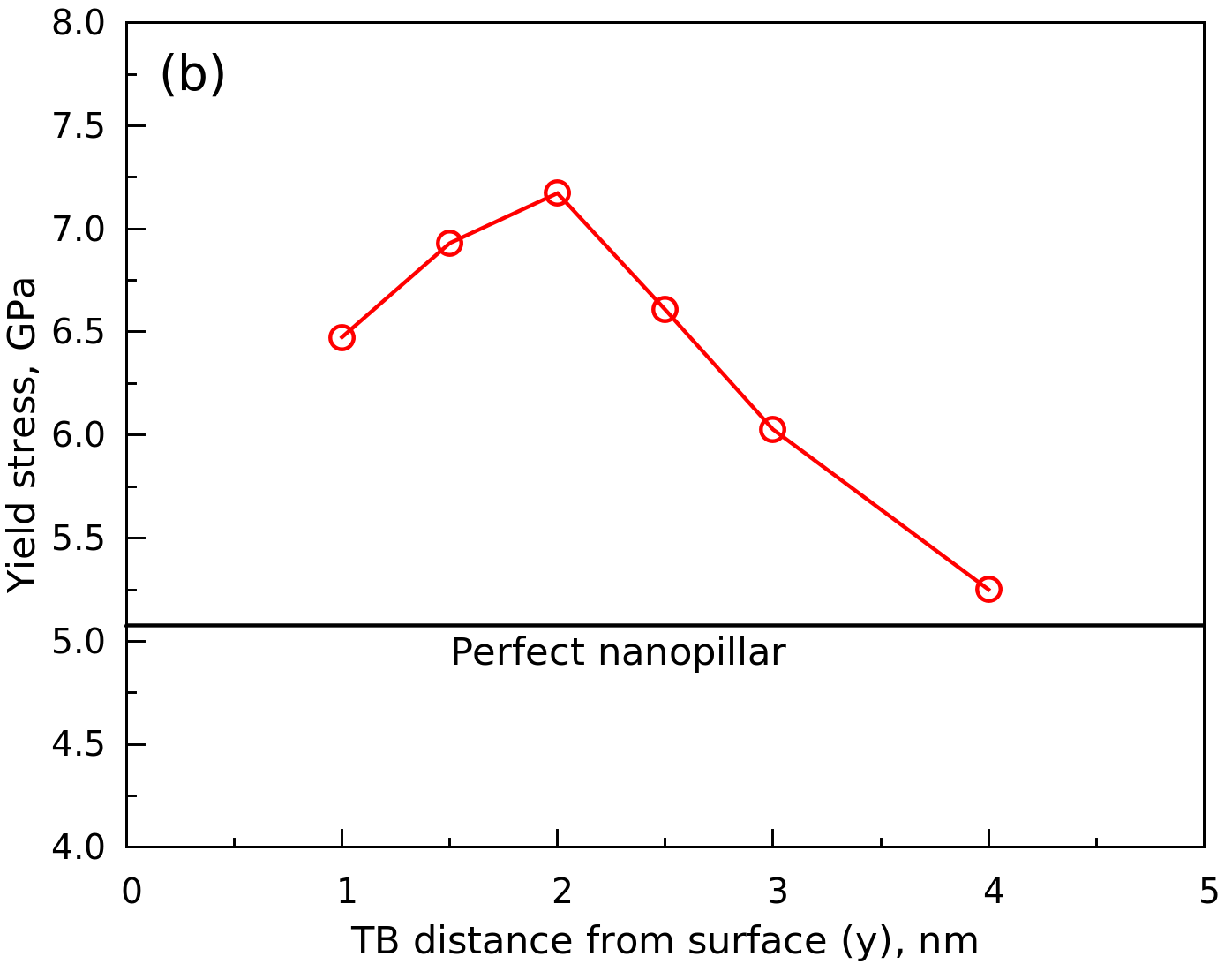}
%  \caption{}
\end{subfigure}
 \caption {(a) Stress-strain behaviour of $<$112$>$ oriented twinned Cu nanopillars with d = 10 nm containing two twin 
 boundaries located at various positions or distances (y) from the nanopillar surface. (b) Variation of yield strength 
 as a function of twin boundary distance (y) from the nanopillar surface. For comparison, the yield strength of twin 
 free $<$112$>$ nanopillar is shown as horizontal line in (b).} 
 \label{2TBs-SS}
 \end{figure}

\begin{figure}[h]
\centering
\begin{subfigure}[b]{0.485\textwidth}
\includegraphics[width=\textwidth]{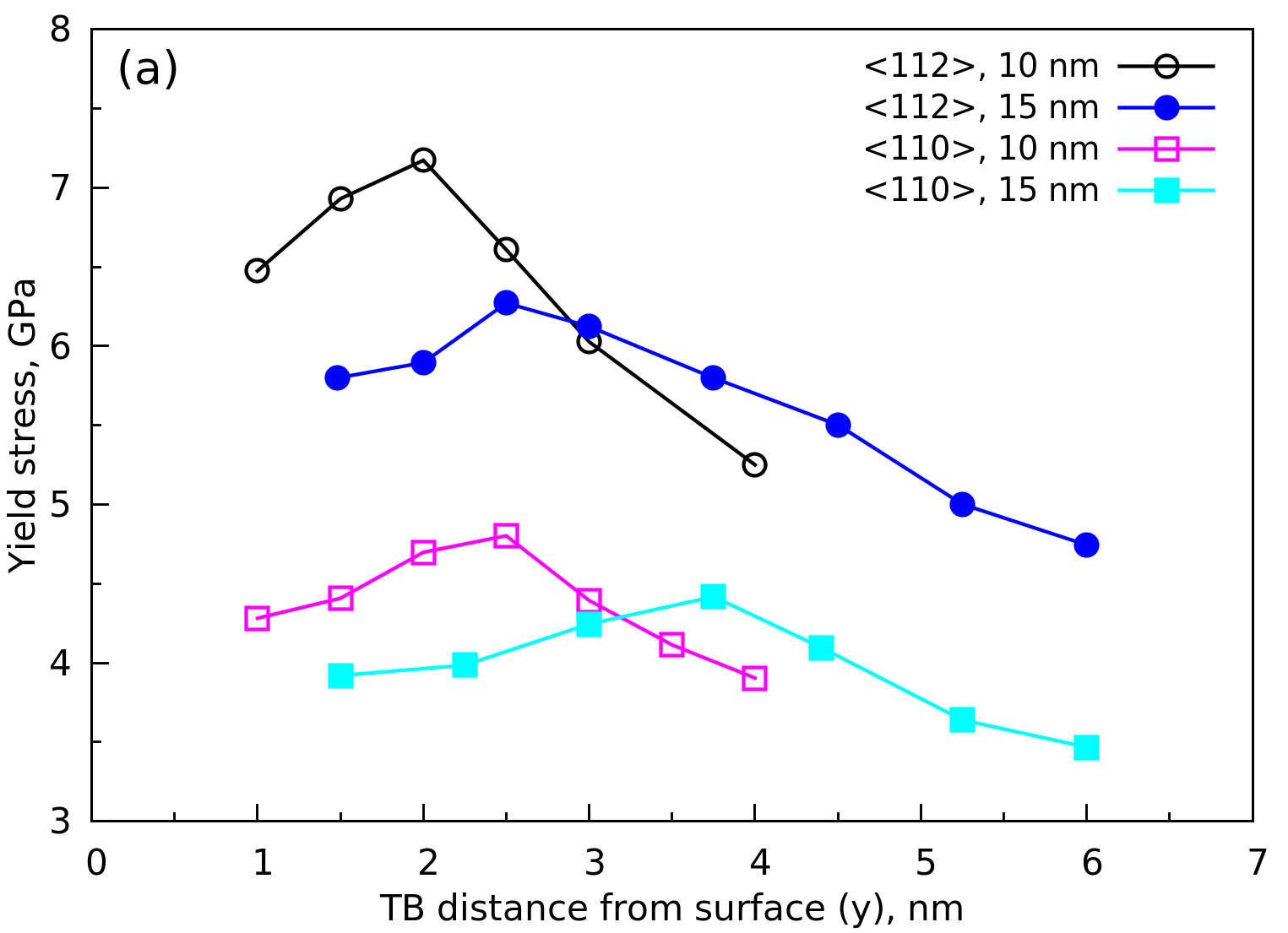}
%                \caption{}
\end{subfigure}
%\qquad
\begin{subfigure}[b]{0.485\textwidth}
\includegraphics[width=\textwidth]{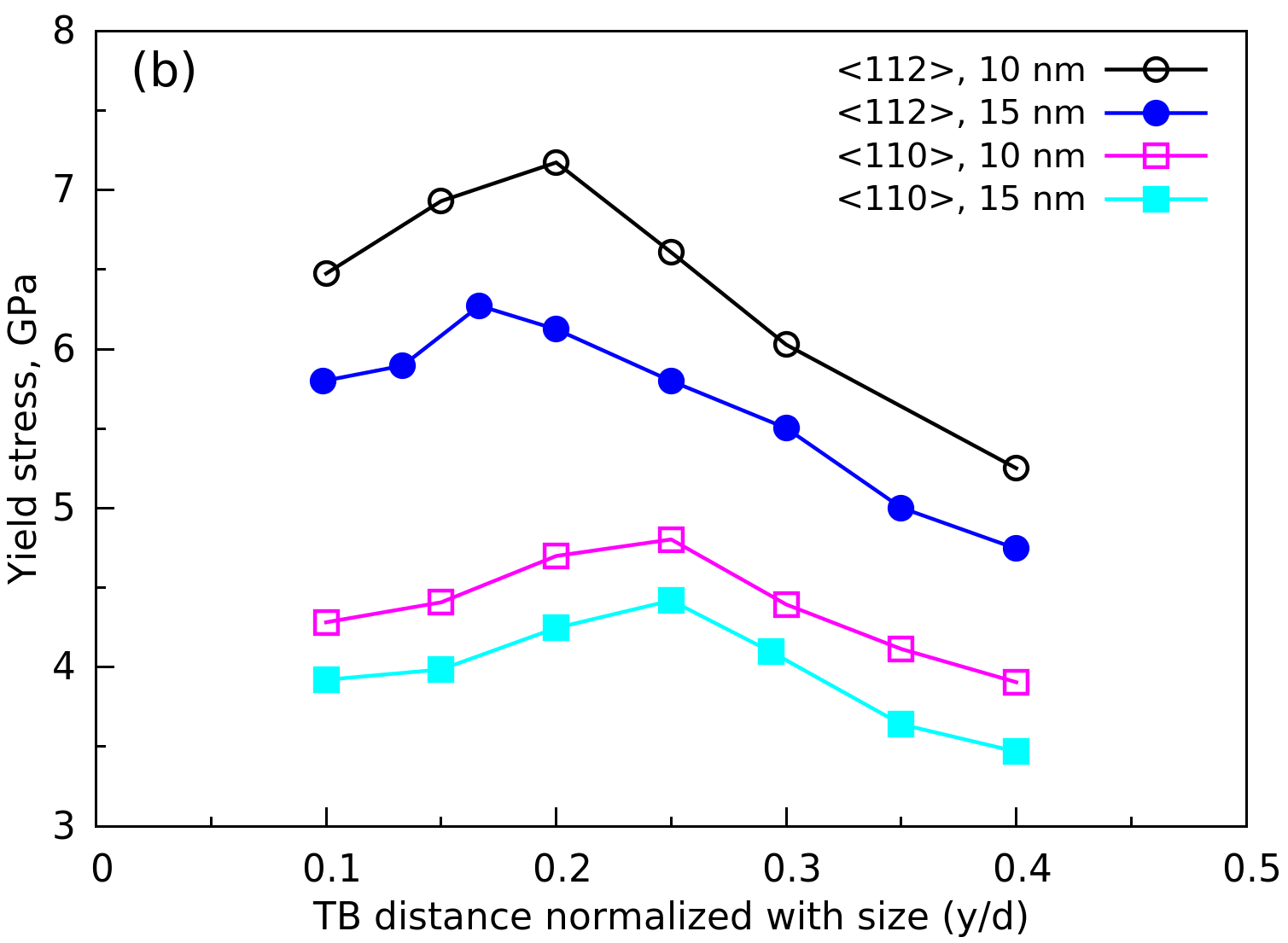}
%  \caption{}
\end{subfigure}
 \caption {The effect of size and orientation on the variation of yield strength as function of \hl{(a) twin boundary distance 'y' from the 
 surface, and (b) normalized twin boundary distance 'y/d' in nanopillar containing two twin boundaries.}} 
 \label{2TBs-YS-Size}
 \end{figure}

In addition to nanopillar containing a single TB, it is also interesting to examine the role of TB position when the nanopillar 
contains multiple TBs. For this, two TBs have been placed in the nanopillar, each at a distance of y from the surface (Figure 
\ref{Initial}b) and y is varied systematically. Figure \ref{2TBs-SS}a shows the tensile stress-strain behaviour of $<$112$>$ Cu 
nanopillars with $d = 10$ nm containing two TBs located at various positions or distance (y) from the surface. It can be seen 
that, overall the stress-strain behaviour is almost similar to that described for $<$112$>$ nanopillar containing a single TB, 
which consist of an initial elastic deformation till the peak followed by a sudden drop in flow stress and then large fluctuations 
consisting of peaks and troughs during plastic deformation. However, as shown in Figure \ref{2TBs-SS}b, the yield stress of 
$<$112$>$ nanopillar containing two TBs displayed a quit different behaviour as a function of TB position. At small distances from the 
pillar surface, the yield stress increases with increasing distance and reaches a maximum value at a distance of y = 2 nm 
followed by gradual decrease. Similar yield stress behaviour has been observed for nanopillar with different orientation 
($<$110$>$) and size (Figure \ref{2TBs-YS-Size}a-b). However, the important difference is the position of TB at which the nanopillar 
attains the peak strength. As can be seen in Figure \ref{2TBs-YS-Size}a, for the same size (d = 10 nm), the peak strength in 
$<$110$>$ nanopillars is observed at marginally higher value of y than that observed for $<$112$>$ nanopillar. Similarly, for 
both $<$112$>$ and $<$110$>$ nanopillars, the position at which the nanopillars attains the peak strength increases with 
increasing size (Figure \ref{2TBs-YS-Size}a). Specifically, in $<$110$>$ nanopillars this position is found to be at y = 2.5 nm 
and 3.75 nm for the pillar 
sizes of 10 and 15 nm, respectively. \hl{For better understanding, the yield strength is plotted as a function of TB distance 
(y) normalized with nanopillar size (d) in Figure \mbox{\ref{2TBs-YS-Size}}b. The results suggest that, irrespective of size, 
the strength is maximum in $<$110$>$ nanopillars when the twin boundaries are placed at a y/d ratio of 0.25, i.e., at a 
distance of one fourth the nanopillar size from the surfaces. In $<$112$>$ nanopillars, the strength is maximum for the y/d 
ratio of approximately 0.2 \mbox{(Figure \ref{2TBs-YS-Size}b)}, i.e., y should be one fifth the nanopillar size}. Thus, 
irrespective of size and orientation, the TB position significantly influences the yield strength of twinned nanopillars.

\begin{figure}[h]
\centering
\includegraphics[width=9cm]{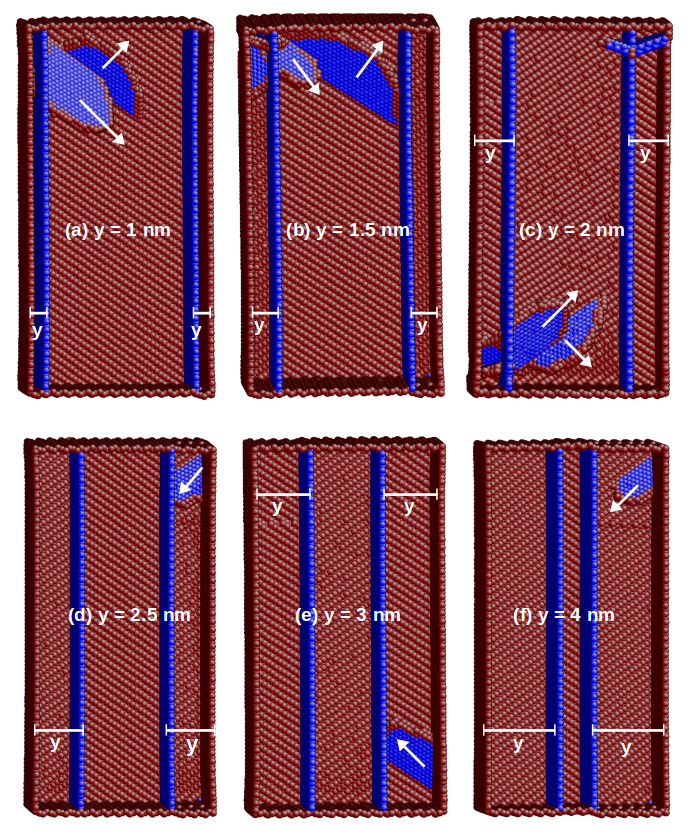}
\caption {The dislocation nucleation during the yielding of $<$112$>$ Cu nanopillar with d = 10 nm containing two twin boundaries 
located at a distance of (a) y = 1 nm, (b) y = 1.5 nm, (c) y = 2 nm, (d) y = 2.5 nm, (e) y = 3 nm, and (f) y = 4 nm from nanopillar 
surface. The atoms are coloured according to their CNA parameter. The front surfaces and the perfect atoms are removed for better 
visualization of dislocations. The blue colour atoms indicate the HCP or stacking fault atoms, while the red colour atoms represent 
the surfaces along with dislocation core atoms.}
\label{2TBs-yielding}
\end{figure}

Unlike the nanopillar containing a single TB, the yield strength of the nanopillar containing two TBs shows a rather different 
behaviour (Figures \ref{yieldstress} and \ref{2TBs-YS-Size}). According to equation \ref{equation1}, the repulsive force should 
decrease with increasing $y$ suggesting that the yield strength should also show a decreasing trend with increasing twin 
boundary distance from the surface. However, this behaviour in yield strength is observed only at larger values of $y$ (Figure 
\ref{2TBs-YS-Size}). At smaller values of $y$, the yield strength increases with increasing the TB distance from the surface. 
In order to understand this anomalous behaviour, the defect nucleation during yielding has been carefully examined. Figure 
\ref{2TBs-yielding} shows the dislocation nucleation in $<$112$>$ nanopillar containing two TBs located at various positions 
$y$ from the surface. It can be seen that, at small distances ($y$ = 1,1.5 and 2 nm), the yielding occurs in the middle grain, 
which is also large in size (Figure \ref{2TBs-yielding}a-c). Specifically, the dislocations nucleate from intersection of TB 
and surface. Contrary to this, the yielding at higher TB distances ($y > 2$ nm) occurs in the grains closer the surface (Figure 
\ref{2TBs-yielding}d-f). Further in this case, no nucleation of dislocations is observed in the middle grain. In other words, 
there is a change in nucleation behaviour from middle grain to surface grains with increasing TB position from nanopillar 
surface. This change in dislocation nucleation behaviour is directly reflected in yield strength behaviour. At small distances 
where nucleation is observed in the middle grain, the repulsive force increases with increasing TB position $y$ as the TBs come 
closer to nucleation points, i.e., for small values of $y$, the repulsive force is inversely proportional to the spacing between 
the twin boundaries ($d-2y$). On the contrary, at large TB distances, where nucleation is observed in the grains near to surface, 
the repulsive force decreases with increasing TB position $y$ as the TBs move away from nucleation points. This results in decrease 
of yield strength with increasing TB distance at large values of $y$. \hl{It can be seen that in nanopillars with single and double 
TBs, the dislocation nucleation occurs mainly in the larger grains (Figures \mbox{\ref{yielding}} and \mbox{\ref{2TBs-yielding})}. 
This may be due to the availability of more volume (for dislocation glide) within the large grain, which decreases the activation 
energy for dislocation nucleation. As a result of this, Deng and Sansoz \mbox{\cite{RF-Sansoz}} have shown that the yield stress 
of twinned nanopillars depends only on the largest distance between the TB and dislocations, i.e., size of the largest grain where 
the dislocations nucleate. This phenomena is analogous to dislocation nucleation with respect to nanowire size, where dislocation 
nucleation is easier in large size nanowires compared to smaller ones \cite{Sainath-PhilMag17}.}

\hl{It is well known that the nanopillar surfaces exerts an image force on dislocation nucleation, thus resulting in size dependent 
yield strength \mbox{\cite{IF1,IF2}}. The present results suggest that due to the repulsive force of TBs, the TB position from the 
nanopillar surface becomes 
an important factor in dictating the yield strength of twinned nanopillars. However, any kind of attractive/repulsive force between the 
TBs and surfaces, and TB-TB is not known, which needs future attention}. Twin boundary position dependent yield strength as observed in 
the present study has great consequence in twinned nanocrysalline materials and also in materials where TB migration is extensively 
observed. In twinned polycrystalline Cu, Marchenko and Zhang \cite{MetTransA} studied the effect of TB position on the mechanical 
properties by placing two TBs in each grain and varying their position with respect to the grain boundary \cite{MetTransA}. Since the 
dislocations nucleate from grain boundaries (GBs), the distance between TB and GB played an important role on dislocation nucleation, 
strength and toughness. It has been reported that the strength and toughness values reach their maximum at an intermediate distance 
between TB and GB, while it decreases when they were far away or close to each other \cite{MetTransA}. This variation of yield strength 
with respect to TB position in a polycrystalline material is similar to that observed in the present study, where yield strength varies 
with respect to TB position from the nanopillar surface. These results demonstrates that the mechanical properties of the twinned 
nanopillars or nanocrystalline materials can be controlled by carefully tailoring the position of the TBs within nanopillar/grain 
without changing their number density, unlike the case of TB spacing. The strengthening by tailoring the TB position has an advantage 
as compared to the strengthening by TB spacing. The strengthening by TB spacing requires the insertion of multiple TBs separated by an 
equal distance and small spacing requires high density of TBs. In contrast, the strengthening by TB position involves limited number of 
TBs (one or two) and only their position needs to be changed with respect to nanopillar surface/GB to get the desired strength. The present 
results may also suggest that in materials where TB migration is extensively observed \cite{PML-TBM}, the change in the position of the 
TB with respect to the surface/GB may significantly influences the flow stress during the plastic deformation.

\section{Conclusions}

In the present study, the role of twin boundary (TB) position on dislocation nucleation and yield strength of twinned 
nanopillars has been investigated using MD simulations. The results indicated that the TB position or distance from the 
surface significantly influences the yield strength of twinned metallic nanopillars. Further, it has been observed that 
the effect of TB position is not always the same. When the nanopillar contains a single TB, the strength increases as 
the TBs move away from surface and attains the maximum when it is located at the center of the nanopillar. Further, in 
nanopillar containing single TB, the dislocations always nucleate from the surface which is far away from the TB. No 
nucleation of dislocations is observed from surface closer to the TB. Due to this, the repulsive force and correspondingly 
the yield strength increases with increasing TB distance x from the nearby surface (or decreasing TB distance from the 
far away surface, d-x). In case of a nanopillar containing two TBs, the yield strength initially increases with increasing 
TB distance and reaches a peak value before showing a gradually decreasing behaviour. In other words, there is an optimal 
distance (y) of TB at which the nanopillar containing two TBs exhibits a peak strength, which is found to be around one 
fourth \hl{and one fifth the nanopillar size for $<$110$>$ and $<$112$>$ orientations, respectively}. In nanopillar containing 
two TBs, the dislocation nucleate in the middle grain when the TBs are close to surface (small y), while nucleation from the 
grains closer to surface has been observed when they are far away from the surface (large y). As a result, the yield strength 
is maximum at an intermediate distance between TBs and surface (y = d/4), while it decreases when the TBs are too close or 
too far from the surface. This study demonstrates that the mechanical properties of the twinned nanopillars can be controlled 
by carefully tailoring the position of the TBs within nanopillar without changing their number density.

\section*{Data availability}

The data that support the findings in this paper are available from the corresponding author on request.

%\section*{References}


\begin{thebibliography}{99}
 
 
%%%%%%%%%%%%%%%%%%%%%%%%%%% Imp. Reviews

\bibitem{Review1-Cai} C.R. Weinberger, W. Cai, Plasticity of metal nanowires, J. Mater. Chem. 22 (2012) 3277-3292. 

\bibitem{Review2-EML} J.Wang, S.X. Mao, Atomistic perspective on in situ nanomechanics, Extr. Mech. Lett. 8 (2016) 127-139.


%%%%%%%%%%%%%%%%%%%%%% Papers on perfect nanopillars

\bibitem{PRL-Samanta} T. Zhu, J. Li, A. Samanta, A. Leach, K. Gall, Temperature and strain rate dependence of surface 
dislocation nucleation, Phys. Rev. Lett. 100 (2008) 025502.

\bibitem{Volkert-APL} B. Roos, B. Kapelle, G. Richter, C.A. Volkert, Surface dislocation nucleation controlled deformation 
of Au nanowires, Appl. Phys. Lett. 105 (2014) 201908.

\bibitem{Cao2008-Shape} A. Cao, E. Ma, Sample shape and temperature strongly influence the yield strength of metallic nanopillars, Acta 
Mater. 56 (2008) 4816-4828.

\bibitem{Xie2015-Srate} H. Xie, F. Yin, T. Yu, G. Lu, Y. Zhang, A new strain-rate-induced deformation mechanism of Cu nanowire: Transition 
from dislocation nucleation to phase transformation, Acta Mater. 85 (2015) 191-198.

\bibitem{Rohith-CCM} P. Rohith, G. Sainath, B.K. Choudhary, Effect of orientation and mode of loading on deformation behaviour 
of Cu nanowires, Comp. Condens. Mat. 17 (2018) e00330.

\bibitem{Sainath-PhilMag17} G. Sainath, P. Rohith, B.K Choudhary, Size dependent deformation behaviour and dislocation 
mechanisms in $<$100$>$ Cu nanowires, Philos. Mag. 97 (2017) 2632-2657.

%%%%%%%%%%%%%%%%%%%%%%%%%%%%%%%% Twinned nanopillars

\bibitem{Cao2007-TBs} A.J. Cao, Y.G. Wei, S.X. Mao, Deformation mechanisms of face-centered-cubic metal nanowires with twin boundaries, 
Appl. Phys. Lett. 90 (2007) 151909.

\bibitem{Sansoz-NanoLett} K.A. Afanasyev, F. Sansoz, Strengthening in gold nanopillars with nanoscale twins, Nano Lett. 
7 (2007) 2056-2062.

\bibitem{Lu2009-Sci} L. Lu, X. Chen, Revealing the maximum strength in nanotwinned copper, Science 323 (2009) 607-610.

\bibitem{Jang2012-Natnano} D. Jang, X. Li, H. Gao, J.R. Greer, Deformation mechanisms in nanotwinned metal nanopillars, Nat. Nanotechnol. 
7 (2012) 594-601.

\bibitem{Rohith-PhilMag19} P. Rohith, G. Sainath, Sunil Goyal, A. Nagesha, V.S. Srinivasan, Role of axial twin boundaries on 
deformation mechanisms in Cu nanopillars, Philos. Mag. 100 (2019) 529-550.

\bibitem{Yang2017-SciRep} Z. Yang, L. Zheng, Y. Yue, Z. Lu, Effects of twin orientation and spacing on the mechanical properties of Cu 
nanowires, Sci. Rep. 7 (2017) 10056. 

\bibitem{Sainath-PhilMag16} G. Sainath, B.K. Choudhary, Deformation behaviour of body centered cubic iron nanopillars 
containing coherent twin boundaries, Philos. Magaz. 96 (2016) 3502-3523.


%%%%%%%%%%%%%%%%%%%%%%%%%%%%%%%%%%%% Repulsive force papers


\bibitem{RF-PRB} Z. Chen, Z. Jin, and H. Gao, Repulsive force between screw dislocation and coherent twin boundary in 
aluminum and copper, Phys. Rev. B 75 (2007) 212104.

\bibitem{RF-Sansoz} C. Deng, F. Sansoz, Repulsive force of twin boundary on curved dislocations and its role on the 
yielding of twinned nanowires, Scr. Mater. 63 (2010) 50-53.

\bibitem{RF-Acta} X. Guo and Y. Xia, Repulsive force vs. source number: Competing mechanisms in the yield of twinned gold 
nanowires of finite length, Acta Mater. 59 (2011) 2350-2357.

%%%%%%%%%%%%%%%%%%%%%%%%%%%%%%%%%%%%%%%% TB Spacing


%%%%%%%%%%%%%%%%%%%%%%%%%%%%%%%%%%%%%%%%%%%%%%5 Longitudinal TBs


\bibitem{SainathPLA} G. Sainath, B.K. Choudhary, Molecular dynamics simulation of twin boundary effect on deformation of Cu 
nanopillars, Phys. Lett. A 379 (2015) 1902-1905.

\bibitem{PRL-TBP} G. Cheng, S. Yin, T.-H. Chang, G. Richter, H. Gao, and Y. Zhu, Anomalous tensile de-twinning in twinned nanowires, 
Phys. Rev. Lett. 119 (2017), p. 256101.

\bibitem{Acta-Axial} W.S. Ko, A. Stukowski, R. Hadian, A. Nematollahi, J.B. Jeon, W.S. Choi, G. Dehm, J. Neugebauer, C. Kirchlechner, 
B. Grabowski, Atomistic deformation behavior of single and twin crystalline Cu nanopillars with preexisting dislocations, 
Acta Mater. 197 (2020) 54-68.

\bibitem{Jeon-Scripta} J.B. Jeon, G. Dehm, Formation of dislocation networks in a coherent Cu $\Sigma3(111)$ twin boundary, 
Scr. Mater. 102 (2015) 71-74.

%%%%%%%%%%%%%%%%%%%%%%%%%%%%%%%%%%%%%%%%%%%%%%%%%% Simulation details


\bibitem{Plimpton-1995} S. Plimpton, Fast parallel algorithms for short-range molecular dynamics, J. Comp. Phy. 117 (1995) 1-19.

\bibitem{Mishin-2001} Y.Mishin, M.J. Mehl, D.A. Papaconstantopoulos, A.F. Voter, J.D. Kress,  Structural stability and lattice 
defects in copper: Ab initio, tight-binding, and embedded-atom calculations, Phys. Rev. B, 63 (2001) 1-16.

\bibitem{Liang-PRB} W. Liang, M. Zhou, Atomistic simulations reveal shape memory of fcc metal nanowires, Phys. Rev. B 73 (2006) 
115409.

\bibitem{Virial} J.A. Zimmerman, E.B. Webb, J.J. Hoyt, R.E. Jones, P.A. Klein, D.J. Bammann,  Calculation of stress in atomistic 
simulations, Modell. Simul. Mater. Sci. Eng. 12 (2003) S319-s332.

\bibitem{AtomEye} J. Li, AtomEye: an efficient atomistic configuration viewer, Modell. Simul. Mater. Sci. Eng. 11 (2003) 173-177.

\bibitem{Ovito} A. Stukowski, Visualization and analysis of atomistic simulation data with OVITO-the Open Visualization Tool, 
Modell. Simul. Mater. Sci. Eng. 18 (2010) 015012.

%%%%%%%%%%%%%%%%%%%%%%%%%%%%%%%%%%%%%%%%%%%%%%%%%%%%%%%% Results and Discussion

\bibitem{descrete} H. Zheng, A. Cao, C.R. Weinberger, J.Y. Huang, K. Du, J. Wang, Y. Ma, Y. Xia and S.X. Mao, Discrete plasticity 
in sub-10-nm-sized gold crystals, Nat Commun. 1 (2010) 144.

\bibitem{Deng2009} C. Deng, F. Sansoz, Size-dependent yield stress in twinned gold nanowires mediated by site-specific surface dislocation 
emission, Appl. Phys. Lett. 95 (2009) 091914.

\bibitem{IF1}W. Ye, A. Ougazzaden, M. Cherkaoui, Analytical formulations of image forces on dislocations with surface stress in nanowires 
and nanorods, Inter. J. Solids and Struct. 50 (2013) 4341-4348.

\bibitem{IF2} Q.-J. Li, B. Xu, S. Hara, J. Li, E. Ma, Sample size-dependent surface dislocation nucleation in nanoscale crystals, 
Acta Mater. 145 (2018) 19-29.

\bibitem{MetTransA} A. Marchenko, H. Zhang, Effects of location of twin boundaries and grain size on plastic deformation of nanocrystalline 
copper, Metall. and Mat. Trans. A 43A (2012) 3547-3555.

\bibitem{PML-TBM} Y.B. Wang, M.L. Sui, and E. Ma, In situ observation of twin boundary migration in copper with nanoscale twins 
during tensile deformation, Phil. Mag. Lett. 87 (2007) 935-942.

\end{thebibliography}
\end{document}